\documentclass[]{article}
\usepackage[T1]{fontenc}
\usepackage{lmodern}
\usepackage{amssymb,amsmath}
\usepackage{ifxetex,ifluatex}
\usepackage{fixltx2e} 
\IfFileExists{upquote.sty}{\usepackage{upquote}}{}
\ifnum 0\ifxetex 1\fi\ifluatex 1\fi=0 
  \usepackage[utf8]{inputenc}
\else 
  \ifxetex
    \usepackage{mathspec}
    \usepackage{xltxtra,xunicode}
  \else
    \usepackage{fontspec}
  \fi
  \defaultfontfeatures{Mapping=tex-text,Scale=MatchLowercase}
  
\fi
\IfFileExists{microtype.sty}{\usepackage{microtype}}{}
\usepackage{color}
\usepackage{fancyvrb}

\DefineVerbatimEnvironment{Highlighting}{Verbatim}{commandchars=\\\{\}}
\newenvironment{Shaded}{}{}

\newcommand{\DecValTok}[1]{\textcolor[rgb]{0.25,0.63,0.44}{{#1}}}

\newcommand{\CommentTok}[1]{\textcolor[rgb]{0.38,0.63,0.69}{\textit{{#1}}}}
\newcommand{\OtherTok}[1]{\textcolor[rgb]{0.00,0.44,0.13}{{#1}}}

\newcommand{\NormalTok}[1]{{#1}}
\usepackage{longtable}
\usepackage{graphicx}
\makeatletter
\def\ScaleIfNeeded{%
  \ifdim\Gin@nat@width>\linewidth
    \linewidth
  \else
    \Gin@nat@width
  \fi
}
\makeatother
\let\Oldincludegraphics\includegraphics
{%
 \catcode`\@=11\relax%
 \gdef\includegraphics{\@ifnextchar[{\Oldincludegraphics}{\Oldincludegraphics[width=\ScaleIfNeeded]}}%
}%
\ifxetex
  \usepackage[setpagesize=false, 
              unicode=false, 
              xetex]{hyperref}
\else
  \usepackage[unicode=true]{hyperref}
\fi
\hypersetup{breaklinks=true,
            bookmarks=true,
            pdfauthor={Mukkai Krishnamoorthy; Kenneth Simons; Benjamin Pringle},
            pdftitle={Case study of approaches to finding patterns in citation networks},
            colorlinks=true,
            citecolor=blue,
            urlcolor=blue,
            linkcolor=magenta,
            pdfborder={0 0 0}}
\title{Case study of approaches to finding patterns in citation networks}
\author{Mukkai Krishnamoorthy; Kenneth Simons; Benjamin Pringle}
\date{}

\begin{document}
\maketitle

\paragraph*{Abstract}\label{abstract}
\addcontentsline{toc}{paragraph}{Abstract}

Analysis of a dataset including a network of LED patents and their
metadata is carried out using several methods in order to answer
questions about the domain. We are interested in finding the
relationship between the metadata and the network structure; for
example, are central patents in the network produced by larger or
smaller companies?

We begin by exploring the structure of the network without any metadata,
applying known techniques in citation analysis and a simple clustering
scheme. These techinques are then combined with metadata analysis to
draw preliminary conclusions about the dataset.

\section{Introduction}\label{introduction}

A \textbf{citation network} is a \textbf{graph} representing citations
between documents such as scholarly articles or patents. Each document
is represented by a \textbf{node} in the graph, and each citation is
represented by an \textbf{edge} connecting the \emph{citing} node to the
\emph{cited} node.

Earlier work in the area of citation network analysis by Garfield, Sher,
and Torpie (1964) popularized the systematic use of forward citation
count as a metric for scholarly influence. Hummon and Dereian (1989)
defined several new metrics to track paths of influence, which were
later improved by Batagelj (2003). The PageRank algorithm was introduced
by Page et al. (1999). It originally powered the Google search engine,
treating hypertext links as ``citations'' between documents on the world
wide web. These are merely a select few prior works -- this listing
fails to exhaust even the highlights.

\subsection{Case study: LED patents}\label{case-study-led-patents}

In this paper, we will be using a network of roughly one hundred
thousand LED patent applications supplied by Simons (2011).

All data is stored in plain \texttt{latin-1}-encoded text, with one row
of data per line of text, and fields separated by tab characters.

Each patent application has a unique identifier: \texttt{applnID}.

The dataset includes a list of all citations (mapping the citing
\texttt{applnID} to the cited \texttt{applnID}), in addition to several
metadata fields:

\begin{itemize}
\itemsep1pt\parskip0pt\parsep0pt
\item
  \texttt{appMyName} -- normalized name of company applying for patent
\end{itemize}

\subsubsection{A very brief history of LED
patents}\label{a-very-brief-history-of-led-patents}

Partridge (1976) filed the first patent demonstrating
electroluminescence from polymer films, one of the key advances that
lead to the development of organic LEDs. (This is \texttt{applnID}
47614741 in our dataset.)

Kodak researchers VanSlyke and Tang (1985) built on this work when they
filed a new patent demonstrating improved power conversion in organic
electroluminescent devices. (This is \texttt{applnID} 51204521 in our
dataset.) Another group of Kodak scientists, Tang, Chen, and Goswami
(1988), patented the first organic LED device, now used in televisions,
monitors, and phones.

This background helps to validate our methods for classifying patents as
``important.'' A good algorithm should classify the 47614741 and
51204521 nodes as significant. When we present our techniques, we will
use this as one metric of success.

\subsection{Computation}\label{computation}

The computation for our analysis was performed using the Python
programming language (\url{http://python.org/}) and the following
libraries:

\begin{itemize}
\itemsep1pt\parskip0pt\parsep0pt
\item
  \texttt{networkx} for network representation and analysis (Hagberg,
  Swart, and S Chult 2008)
\item
  \texttt{pandas} for tabular data analysis (McKinney 2012)
\item
  \texttt{scipy} for statistics (Jones et al. 2001)
\item
  \texttt{matplotlib} for creating plots (Hunter 2007)
\end{itemize}

More information about the code written for this paper can be found
under the section, \hyperref[code]{Code}.

\section{Approaches}\label{approaches}

\subsection{Network structure}\label{network-structure}

The graph has 127,526 nodes and 327,479 edges.

\subsubsection{Forward citations
(indegree)}\label{forward-citations-indegree}

\begin{figure}[htbp]
\centering
\includegraphics{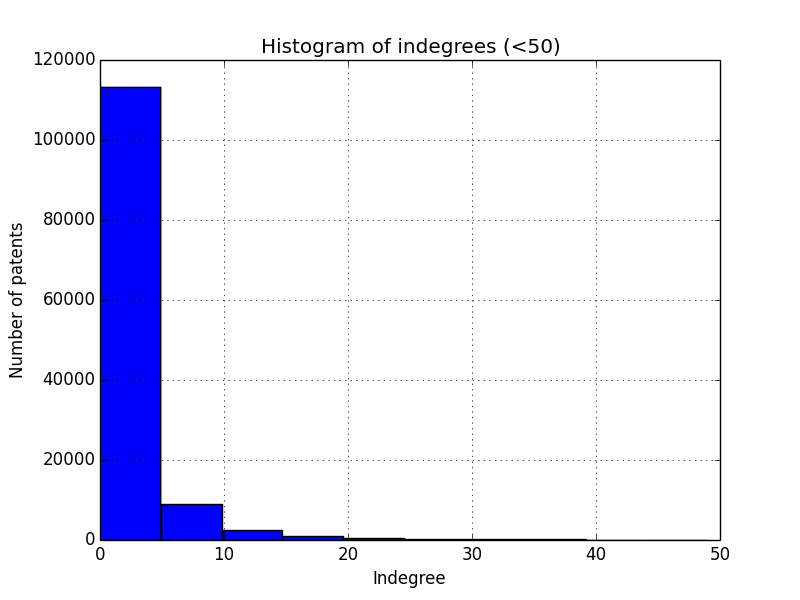}
\caption{Histogram of patents with under 50 citations}
\end{figure}

\begin{figure}[htbp]
\centering
\includegraphics{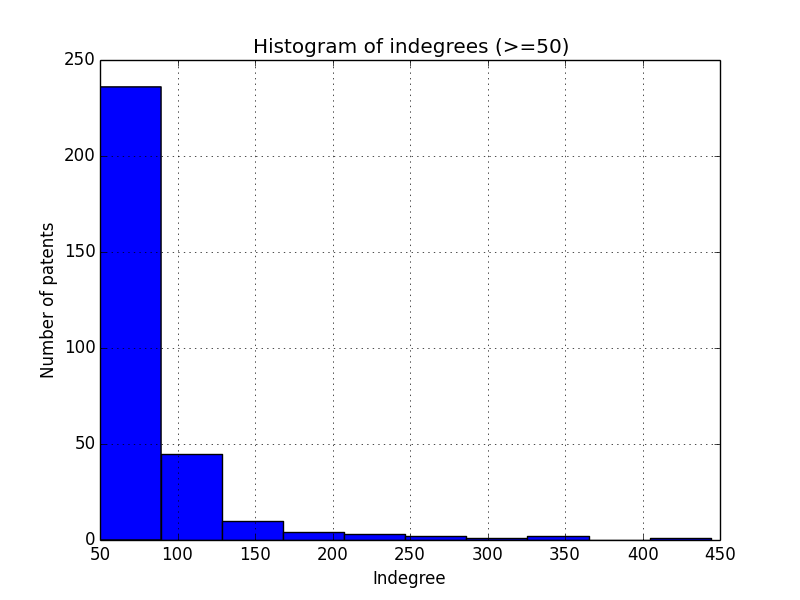}
\caption{Histogram of patents with 50 or more citations}
\end{figure}

Popularized by Garfield, Sher, and Torpie (1964), the simplest way to
determine a patent's relative importance is counting its forward
citations -- that is, other patents which cite the patent in question.
In a citation network where edges are drawn from the citing patent to
the cited patent, the number of forward citations for a given node is
its \textbf{indegree}, or the number of edges ending at the given node.

In our data, 89\% of patents have fewer than 5 citations, and 99\% have
fewer than 50. Nevertheless, there is a small group of slightly over
fifty patents with at least a hundred citations each.

The top ten most-cited patents in our dataset are shown in a table
below:

\begin{longtable}[c]{@{}rr@{}}
\hline\noalign{\medskip}
applnID & indegree
\\\noalign{\medskip}
\hline\noalign{\medskip}
47614741 & 444
\\\noalign{\medskip}
51204521 & 360
\\\noalign{\medskip}
52376694 & 339
\\\noalign{\medskip}
48351911 & 305
\\\noalign{\medskip}
45787627 & 283
\\\noalign{\medskip}
45787665 & 267
\\\noalign{\medskip}
46666643 & 235
\\\noalign{\medskip}
53608703 & 213
\\\noalign{\medskip}
54068562 & 213
\\\noalign{\medskip}
23000850 & 203
\\\noalign{\medskip}
\hline
\end{longtable}

\paragraph{Computation}\label{computation-1}

We computed indegree using \texttt{networkx.DiGraph.in\_degree()}
(Hagberg, Swart, and S Chult 2008).

\subsubsection{PageRank}\label{pagerank}

Another technique for classifying important nodes in a graph is PageRank
(Page et al. 1999), a famous algorithm used by the Google search engine
to rank web pages.

PageRank calculates the probability that someone randomly following
citations will arrive at a given patent. The damping factor $d$
represents the probability at each step that the reader will continue on
to the next patent.

For each patent in our dataset, we calculated:

\begin{itemize}
\itemsep1pt\parskip0pt\parsep0pt
\item
  \texttt{pagescore} -- raw PageRank score (probability 0 to 1)
\item
  \texttt{page\_rank} -- relative numerical rank of the patent (by
  PageRank)
\item
  \texttt{indegree} -- number of forward citations
\item
  \texttt{indegree\_rank} -- relative numerical rank of the patent (by
  indegree)
\end{itemize}

The following chart shows the top ten patents sorted by PageRank:

\begin{longtable}[c]{@{}rllll@{}}
\hline\noalign{\medskip}
applnID & pagescore & page\_rank & indegree & indegree\_rank
\\\noalign{\medskip}
\hline\noalign{\medskip}
47614741 & 0.000371 & 1 & 444 & 1
\\\noalign{\medskip}
51204521 & 0.000329 & 2 & 360 & 2
\\\noalign{\medskip}
48351911 & 0.000291 & 3 & 305 & 4
\\\noalign{\medskip}
45787627 & 0.000241 & 4 & 283 & 5
\\\noalign{\medskip}
48112868 & 0.000227 & 5 & 63 & 172
\\\noalign{\medskip}
45787665 & 0.000220 & 6 & 267 & 6
\\\noalign{\medskip}
52376694 & 0.000210 & 7 & 339 & 3
\\\noalign{\medskip}
53608703 & 0.000193 & 8 & 213 & 8
\\\noalign{\medskip}
46666643 & 0.000173 & 9 & 235 & 7
\\\noalign{\medskip}
47823143 & 0.000168 & 10 & 47 & 342
\\\noalign{\medskip}
\hline
\end{longtable}

Within our dataset, PageRank and indegree are correlated with a Pearson
product-moment coefficient of $r=.80$.

\paragraph{Computation}\label{computation-2}

We computed PageRank using \texttt{networkx.pagerank\_scipy()} with
\texttt{max\_iter} set to 200 and a damping factor of $d=.85$ (Page et
al. 1999; Hagberg, Swart, and S Chult 2008).

\subsection{Clustering}\label{clustering}

\begin{figure}[htbp]
\centering
\includegraphics{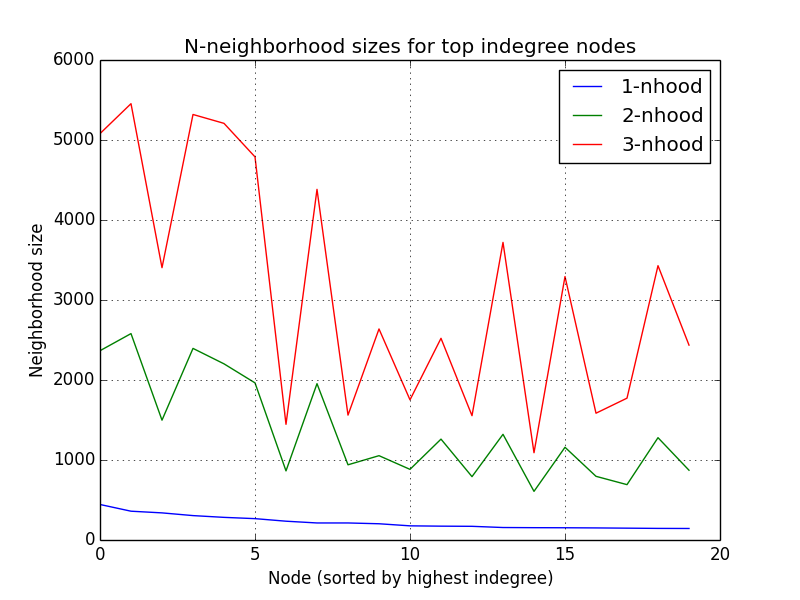}
\caption{Neighborhood sizes for top 20 cited patents}
\end{figure}

As noted by Satuluri and Parthasarathy (2011), most clustering
techniques deal with undirected graphs. We introduce a very simple
technique for defining overlapping clusters in a \emph{directed}
citation network:

\begin{itemize}
\itemsep1pt\parskip0pt\parsep0pt
\item
  Select a small number of highly cited patents as seeds.
\item
  Each seed patent defines a cluster: all patents citing the seed are
  members (its open 1-neighborhood).
\end{itemize}

We considered using larger neighborhoods. The $n$-neighborhood can be
computed recursively by adding all patents citing any patents in the
$(n-1)$-neighborhood. However, these larger neighborhoods grow in size
very quickly. For our purposes of quick computation and visualization,
we chose to keep the smaller clusters from 1-neighborhoods.

This technique creates \emph{overlapping} clusters, where a node can
belong to more than one cluster. Looking at the clusters created from
the top 10 most-cited patents, we computed two measures of overlapping:

\begin{itemize}
\itemsep1pt\parskip0pt\parsep0pt
\item
  \texttt{percentunique} is the fraction of nodes in \emph{only} that
  cluster
\item
  \texttt{bignodes} is the number of seed nodes that appear in the
  cluster (for example, the second cluster contains the seed patent used
  to generate the first cluster, along with three others from our
  original ten seeds)
\end{itemize}

The following chart shows the value of \texttt{percentunique} and
\texttt{bignodes} for each of the ten clusters:

\begin{longtable}[c]{@{}rrr@{}}
\hline\noalign{\medskip}
clustersize & percentunique & bignodes
\\\noalign{\medskip}
\hline\noalign{\medskip}
444 & 0.202703 & 0
\\\noalign{\medskip}
360 & 0.100000 & 4
\\\noalign{\medskip}
339 & 0.280236 & 0
\\\noalign{\medskip}
305 & 0.163934 & 4
\\\noalign{\medskip}
283 & 0.141343 & 0
\\\noalign{\medskip}
267 & 0.101124 & 0
\\\noalign{\medskip}
235 & 0.940426 & 0
\\\noalign{\medskip}
213 & 0.985915 & 0
\\\noalign{\medskip}
213 & 0.464789 & 0
\\\noalign{\medskip}
203 & 0.226601 & 0
\\\noalign{\medskip}
\hline
\end{longtable}

Looking at \texttt{percentunique}, many clusters have a good deal over
overlap, with unique contributions as low as 10\%, although others are
up to 98\% unique. Our analysis will therefore \textbf{not} assume that
these clusters strictly partition the data, and rather look at the
clusters as distinct but potentially overlapping areas of patents.

\begin{figure}[htbp]
\centering
\includegraphics{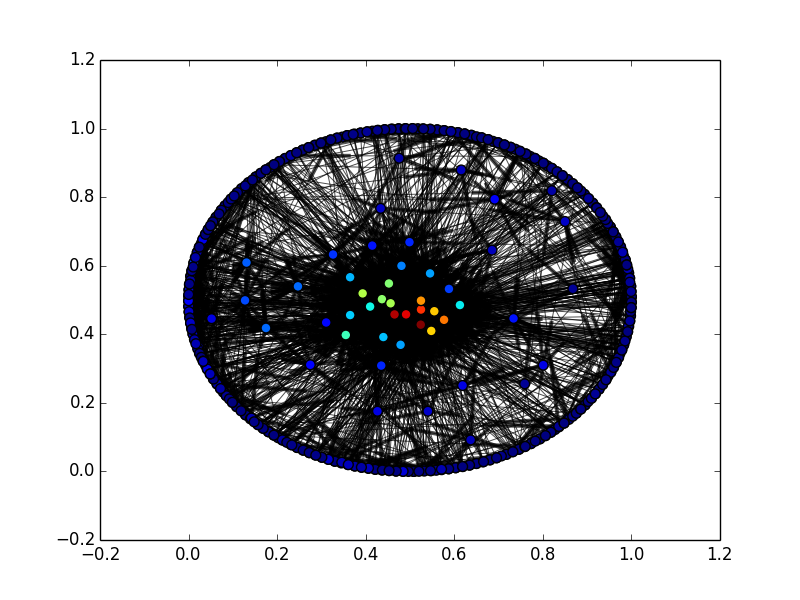}
\caption{1-neighborhood of applnID=47614741 (444 nodes)}
\end{figure}

\begin{figure}[htbp]
\centering
\includegraphics{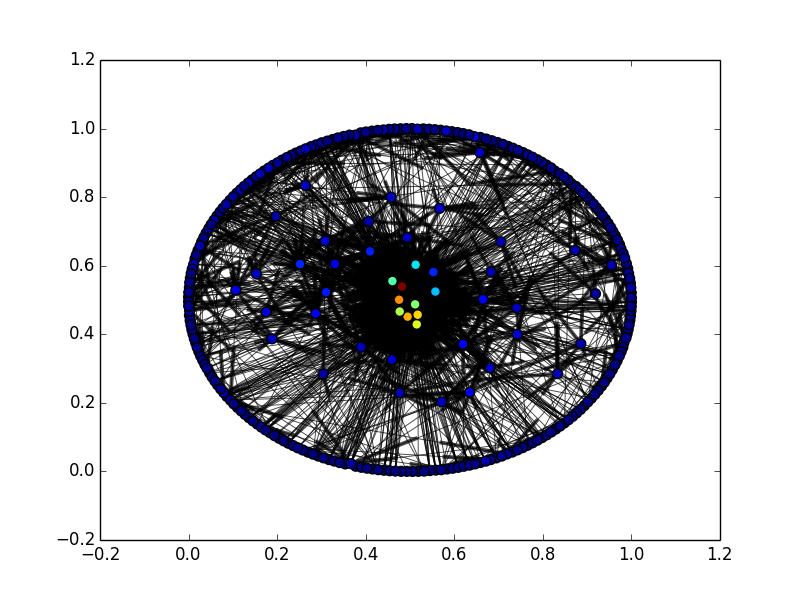}
\caption{1-neighborhood of applnID=45787627 (283 nodes)}
\end{figure}

\begin{figure}[htbp]
\centering
\includegraphics{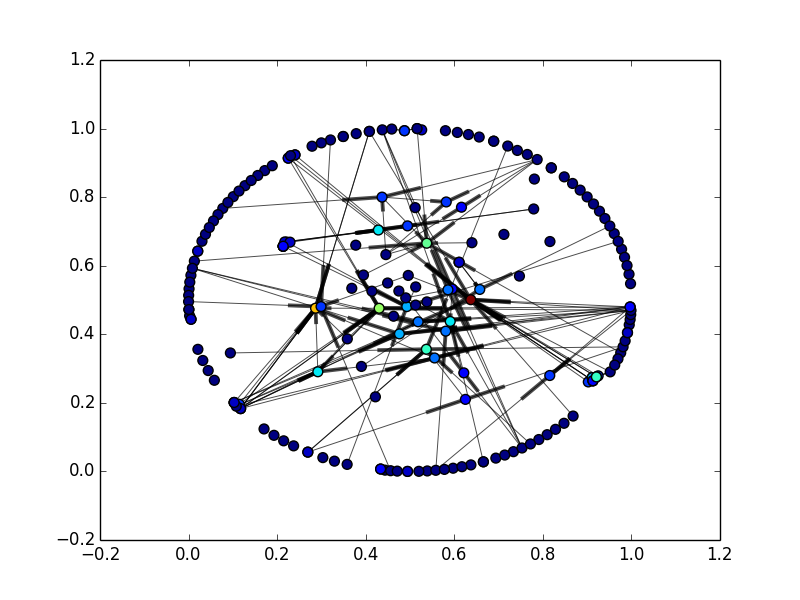}
\caption{1-neighborhood of applnID=23000850 (203 nodes)}
\end{figure}

\paragraph{Computation}\label{computation-3}

The $n$-neighborhood of a node can be computed using the
\hyperref[code]{included code}:

\begin{Shaded}
\begin{Highlighting}[]
\NormalTok{neighborhood(graph, nbunch, depth=}\DecValTok{1}\NormalTok{, closed=}\OtherTok{False}\NormalTok{)}
\end{Highlighting}
\end{Shaded}

\begin{itemize}
\itemsep1pt\parskip0pt\parsep0pt
\item
  \texttt{graph} -- a \texttt{networkx.DiGraph} (see Hagberg, Swart, and
  S Chult 2008)
\item
  \texttt{nbunch} -- a node or nodes in \texttt{graph}
\item
  \texttt{depth} -- the number of iterations (defaults to
  1-neighborhood)
\item
  \texttt{closed} -- set to \texttt{True} if the neighborhood should
  include the root
\end{itemize}

Returns a \texttt{set} containing the neighborhood of the node, or a
\texttt{dict} matching nodes to neighborhood \texttt{set}s.

\subsection{Metadata analytics}\label{metadata-analytics}

Note that only about 35\% of the patents in our dataset (44356 out of
127526) were supplied with \texttt{appMyName} (company name).

\subsubsection{Choosing a metric for company
size}\label{choosing-a-metric-for-company-size}

We would like to explore whether company size has any correlation with
patent quality. Do major innovations originate from big labs, or do
smaller companies pave the way (only to be later acquired)?

In order to begin this investigation, we need a solid metric to quantify
``company size.'' Our first thought was to use a metadata-based
solution, such as the company's net worth or number of employees.
However, it wasn't clear at \emph{which point in time} to measure the
company size -- does a company's employee count in 2013 affect the
quality of a patent it filed in the 1980s?

Instead, we choose a simple metric contained within our dataset: company
size is defined as the \textbf{number of patents submitted}.

This may not be a perfect representation of ``size,'' but it still
allows us to analyze whether these ``prolific'' companies are
contributing any \emph{important} patents or merely a large volume of
consequential patents.

Our set of ``large companies'' will therefore be the 25 companies that
applied for the largest number of patents. They are, in order with
number of LED patents each:

\begin{quote}
\texttt{samsung} (1673), \texttt{semiconductor energy lab} (1437),
\texttt{seiko} (1394), \texttt{sharp} (1103), \texttt{panasonic} (1094),
\texttt{sony} (937), \texttt{toshiba} (848),
\texttt{sanyo (tokyo sanyo electric)} (793), \texttt{philips} (789),
\texttt{kodak} (767), \texttt{hitachi} (632), \texttt{osram} (631),
\texttt{nec} (621), \texttt{lg} (613), \texttt{idemitsu kosan co} (553),
\texttt{canon} (538), \texttt{pioneer} (525), \texttt{mitsubishi} (501),
\texttt{rohm} (420), \texttt{tdk} (384), \texttt{nichia} (370),
\texttt{fujifilm} (369), \texttt{ge} (363), \texttt{sumitomo} (323),
\texttt{lg/philips} (293)
\end{quote}

\subsubsection{Summed outdegree}\label{summed-outdegree}

The ``summed score'' metric isn't very useful in this situation, since
we've already ranked our patents by frequency in our definition of
company size. The summed score for outdegree gives us little new
information.

Below is our list of top 25 patents, with their relative ranking by
summed outdegree score in parentheses:

\begin{quote}
\texttt{samsung} (2), \texttt{semiconductor energy lab} (1),
\texttt{seiko} (3), \texttt{sharp} (5), \texttt{panasonic} (6),
\texttt{sony} (7), \texttt{toshiba} (8),
\texttt{sanyo (tokyo sanyo electric)} (10), \texttt{philips} (9),
\texttt{kodak} (4), \texttt{hitachi} (15), \texttt{osram} (14),
\texttt{nec} (11), \texttt{lg} (17), \texttt{idemitsu kosan co} (12),
\texttt{canon} (16), \texttt{pioneer} (13), \texttt{mitsubishi} (18),
\texttt{rohm} (22), \texttt{tdk} (20), \texttt{nichia} (19),
\texttt{fujifilm} (25), \texttt{ge} (21), \texttt{sumitomo} (26),
\texttt{lg/philips} (27)
\end{quote}

As expected, our top-frequency companies have very high rankings by
summed outdegree score.

\subsubsection{Normalized summed
outdegree}\label{normalized-summed-outdegree}

Instead, we can look at the \emph{normalized} outdegree, or the mean
outdegree of a patent produced by one of our companies. Let's take a
look at just our top 10 companies:

\begin{enumerate}
\def\labelenumi{\arabic{enumi}.}
\itemsep1pt\parskip0pt\parsep0pt
\item
  \texttt{samsung} -- 11.51
\item
  \texttt{semiconductor energy lab} -- 14.91
\item
  \texttt{seiko} -- 13.06
\item
  \texttt{sharp} -- 13.39
\item
  \texttt{panasonic} -- 13.13
\item
  \texttt{sony} -- 13.23
\item
  \texttt{toshiba} -- 14.22
\item
  \texttt{sanyo (tokyo sanyo electric)} -- 13.86
\item
  \texttt{philips} -- 14.47
\item
  \texttt{kodak} -- 19.98
\end{enumerate}

By comparison, the mean outdegree over \emph{all} patents is 5.60.

\subsubsection{Contribution factor --
outdegree}\label{contribution-factor-outdegree}

Let us define patents as relatively significant if their outdegree is in
the 75th percentile. (For our LED dataset, this includes all patents
with at least 11 citations.)

Then, we can calculate contribution factors for each company by finding
the fraction of their patents that are considered relatively
significant. Here are the results:

\begin{enumerate}
\def\labelenumi{\arabic{enumi}.}
\itemsep1pt\parskip0pt\parsep0pt
\item
  \texttt{samsung} -- .63
\item
  \texttt{semiconductor energy lab} -- .85
\item
  \texttt{seiko} -- .78
\item
  \texttt{sharp} -- .86
\item
  \texttt{panasonic} -- .85
\item
  \texttt{sony} -- .82
\item
  \texttt{toshiba} -- .89
\item
  \texttt{sanyo (tokyo sanyo electric)} -- .88
\item
  \texttt{philips} -- .76
\item
  \texttt{kodak} -- .84
\end{enumerate}

\subsubsection{Date partitioning}\label{date-partitioning}

Another interesting approach is to look at the filing date of the
patents. Below is a histogram of number of patents by filing date.

\begin{verbatim}
date range                  count
1940-11-12 to 1945-07-06    1
1945-07-06 to 1950-02-28    6
1950-02-28 to 1954-10-23    107
1954-10-23 to 1959-06-17    247
1959-06-17 to 1964-02-09    369
1964-02-09 to 1968-10-03    344
1968-10-03 to 1973-05-28    362
1973-05-28 to 1978-01-20    575
1978-01-20 to 1982-09-14    678
1982-09-14 to 1987-05-09    1125
1987-05-09 to 1992-01-01    2257
1992-01-01 to 1996-08-25    3451
1996-08-25 to 2001-04-19    8103
2001-04-19 to 2005-12-12    16019
2005-12-12 to 2010-08-06    5040
\end{verbatim}

We can partition each company's patents into thirds -- that is,
\texttt{samsung0} contains the first chronological third of Samsung's
patents, \texttt{samsung1} contains the second third, and
\texttt{samsung2} contains the final third.

We can calculate normalized outdegree for each third:

\begin{verbatim}
company    partition  start       end         normalizedoutdeg    count  totalcount
samsung    0          1989-05-30  2004-06-28  2.6858168761220824  557    1673
samsung    1          2004-06-28  2005-11-30  1.3375224416517055  557    1673
samsung    2          2005-12-02  2010-07-13  0.5116279069767442  559    1673
sel        0          1982-02-09  2002-02-26  8.187891440501044   479    1437
sel        1          2002-02-28  2004-06-23  5.1941544885177455  479    1437
sel        2          2004-06-25  2010-01-06  1.3528183716075157  479    1437
seiko      0          1973-07-13  2002-02-22  5.644396551724138   464    1394
seiko      1          2002-02-25  2004-01-21  2.543103448275862   464    1394
seiko      2          2004-01-21  2009-06-18  0.9978540772532188  466    1394
sharp      0          1972-07-31  1994-02-22  4.809264305177112   367    1103
sharp      1          1994-02-25  2001-10-29  3.5476839237057223  367    1103
sharp      2          2001-10-31  2010-02-26  1.8130081300813008  369    1103
panasonic  0          1963-11-18  1997-10-31  3.4148351648351647  364    1094
panasonic  1          1997-11-05  2002-02-21  3.6950549450549453  364    1094
panasonic  2          2002-02-27  2010-03-05  2.2868852459016393  366    1094
sony       0          1970-04-13  2000-09-11  4.064102564102564   312    937
sony       1          2000-09-14  2003-08-20  4.0576923076923075  312    937
sony       2          2003-08-28  2010-02-10  1.5878594249201279  313    937
toshiba    0          1969-08-25  1993-03-30  4.184397163120567   282    848
toshiba    1          1993-04-13  2001-04-27  6.1063829787234045  282    848
toshiba    2          2001-04-27  2010-03-23  2.3732394366197185  284    848
sanyo      0          1976-12-09  2000-03-17  6.943181818181818   264    793
sanyo      1          2000-03-17  2003-03-28  3.25                264    793
sanyo      2          2003-03-28  2009-01-15  1.4037735849056603  265    793
philips    0          1954-01-29  1999-09-08  6.011406844106464   263    789
philips    1          1999-09-08  2004-07-01  6.068441064638783   263    789
philips    2          2004-07-09  2009-06-03  1.326996197718631   263    789
kodak      0          1965-03-25  2001-01-30  23.63529411764706   255    767
kodak      1          2001-02-02  2003-09-23  4.670588235294118   255    767
kodak      2          2003-09-24  2008-02-25  1.7042801556420233  257    767
\end{verbatim}

\section{Conclusions}\label{conclusions}

Based on our meta-metrics, it appears that while large companies file
many patent applications, these patents are \emph{not} of any lower
quality than average.

By the normalized summed outdegree measure, the top 10 companies each
had a mean outdegree more than \emph{double} that of the entire dataset.

By contribution factor analysis, each of the top 10 (except Samsung)
still exceeded the expected ratio.

\hyperdef{}{code}{\section{Code}\label{code}}

The code for this paper will be posted to GitHub.

Each figure and chart was generated by a different function:

\begin{Shaded}
\begin{Highlighting}[]
\CommentTok{# Analyses}
\NormalTok{big_companies(graph, metadata, show_table=}\OtherTok{True}\NormalTok{)}
\NormalTok{visualize_cluster(graph, index=}\DecValTok{1}\NormalTok{, show_plot=}\OtherTok{False}\NormalTok{)}
\NormalTok{visualize_cluster(graph, index=}\DecValTok{5}\NormalTok{, show_plot=}\OtherTok{False}\NormalTok{)}
\NormalTok{analyze_pagerank(graph, show_table=}\OtherTok{False}\NormalTok{, show_plot=}\OtherTok{False}\NormalTok{)}
\NormalTok{analyze_indegree(graph, show_table=}\OtherTok{False}\NormalTok{, show_plot=}\OtherTok{False}\NormalTok{)}
\NormalTok{analyze_nhood_overlap(graph, show_table=}\OtherTok{False}\NormalTok{)}
\NormalTok{analyze_nhood_size(graph, show_table=}\OtherTok{False}\NormalTok{, show_plot=}\OtherTok{False}\NormalTok{)}
\end{Highlighting}
\end{Shaded}

\section{References}\label{references}

Batagelj, Vladimir. 2003. ``Efficient Algorithms for Citation Network
Analysis.'' \emph{ArXiv Preprint Cs/0309023}.

Garfield, Eugene, Irving H. Sher, and Richard J. Torpie. 1964. ``The Use
of Citation Data in Writing the History of Science.'' DTIC Document.

Hagberg, Aric, Pieter Swart, and Daniel S Chult. 2008. ``Exploring
Network Structure, Dynamics, and Function Using NetworkX.'' Los Alamos
National Laboratory (LANL).

Hummon, Norman P., and Patrick Dereian. 1989. ``Connectivity in a
Citation Network: The Development of DNA Theory.'' \emph{Social
Networks} 11 (1): 39--63.

Hunter, J. D. 2007. ``Matplotlib: A 2D Graphics Environment.''
\emph{Computing In Science \& Engineering} 9 (3): 90--95.

Jones, Eric, Travis Oliphant, Pearu Peterson, and others. 2001. ``SciPy:
Open Source Scientific Tools for Python.'' \url{http://www.scipy.org/}.

McKinney, Wes. 2012. \emph{Python for Data Analysis}. O'Reilly Media.

Page, Lawrence, Sergey Brin, Rajeev Motwani, and Terry Winograd. 1999.
``The PageRank Citation Ranking: Bringing Order to the Web.''

Partridge, Roger Hugh. 1976. ``Radiation Sources.'' Google Patents.

Satuluri, Venu, and Srinivasan Parthasarathy. 2011. ``Symmetrizations
for Clustering Directed Graphs.'' In \emph{Proceedings of the 14th
International Conference on Extending Database Technology}, 343--354.
ACM.

Simons, Kenneth. 2011. ``Files with Patent Data on Technology
Classifications.''

Tang, Ching W., Chin H. Chen, and Ramanuj Goswami. 1988.
``Electroluminescent Device with Modified Thin Film Luminescent Zone.''
Google Patents.

VanSlyke, Steven A., and Ching W. Tang. 1985. ``Organic
Electroluminescent Devices Having Improved Power Conversion
Efficiencies.'' Google Patents.

\end{document}